\documentclass[
    ,final            
  ]
  {aipproc}

\layoutstyle{6x9}

\newcommand{\um}{\,$\mu$m}

\newcommand{\obj}{NGC\,4261}

\newcommand{\aap}{A\&A}

\newcommand{\mnras}{MNRAS}

\begin{document}

\title{Infrared emission in radio galaxy NGC\,4261}

\author{I.M. van Bemmel}{
  address={ESA fellow, Space Telescope Science Institute, Baltimore MD, USA\footnote{Space Telescope Division of ESA}}
}

\author{C.P. Dullemond}{
  address={Max Planck Institute f\''ur Astronomie, Heidelberg, Germany}
}

\author{M. Chiaberge}{
  address={Istituto di Radioastronomia, Bologna, Italy}
}

\author{F.D. Macchetto}{address={Space Telescope Science Institute, Baltimore MD, USA\footnotemark[\value{footnote}]}}

\begin{abstract}
We have analyzed the total and nuclear SED for \obj\ and find that the 
dominant process for the mid- and far-infrared emission in this object is 
non-thermal emission from the active nucleus. Modeling the emission from 
the optically detected 300\,pc dust disk yields no significant disk 
contribution at any wavelength. To explain the observations, either the 
disk has an inflated inner region which partly absorbs the core, or the 
intrinsic core spectrum is curved. The inner 10\,pc of the disk can 
potentially be conceived as an obscuring torus, albeit with optical depth 
around unity. 
\end{abstract}

\maketitle


\section{Introduction}
Unification models for radio-loud active galaxies postulate that
for the weaker FR\,Is only the relativistic jet determines the
appearance of their nuclei, while in pwerful FR\,IIs a circumnuclear 
dust torus adds to the anisotropy \citep{pdb89,up95}. The infrared 
emission from this torus has been used to test unification
models, on the assumption that it dominates the infrared output
of radio galaxies.

We have assessed the contribution of the observed nuclear disk to the
total infrared emission in FR\,I radio galaxy \obj\ (a.k.a. 3C\,270). 
This disk is flatter and has lower optical depth than a typical
obscuring torus, but might be responsible for the absence of a
distinct UV core. We have used 
observations from literature and archives, paying special attention 
to the infrared data. When possible, we measured both total and
unresolved core flux. Between 10\um\ and 1\,mm \obj\ is unresolved.
We will use {\em nuclear disk} for the disk observed in \obj\ and 
{\em torus} for the smaller and denser torus in FR\,IIs. We adopt 
$F_{\nu} \propto \nu^{\alpha}$, $H_0=65$\,km\,s$^{-1}$\,Mpc$^{-1}$ 
and $v_{\rm sys}=2238$\,kms.

\begin{figure*}[!Ht]
\resizebox{8.cm}{!}{\hspace{-1.5cm}\includegraphics{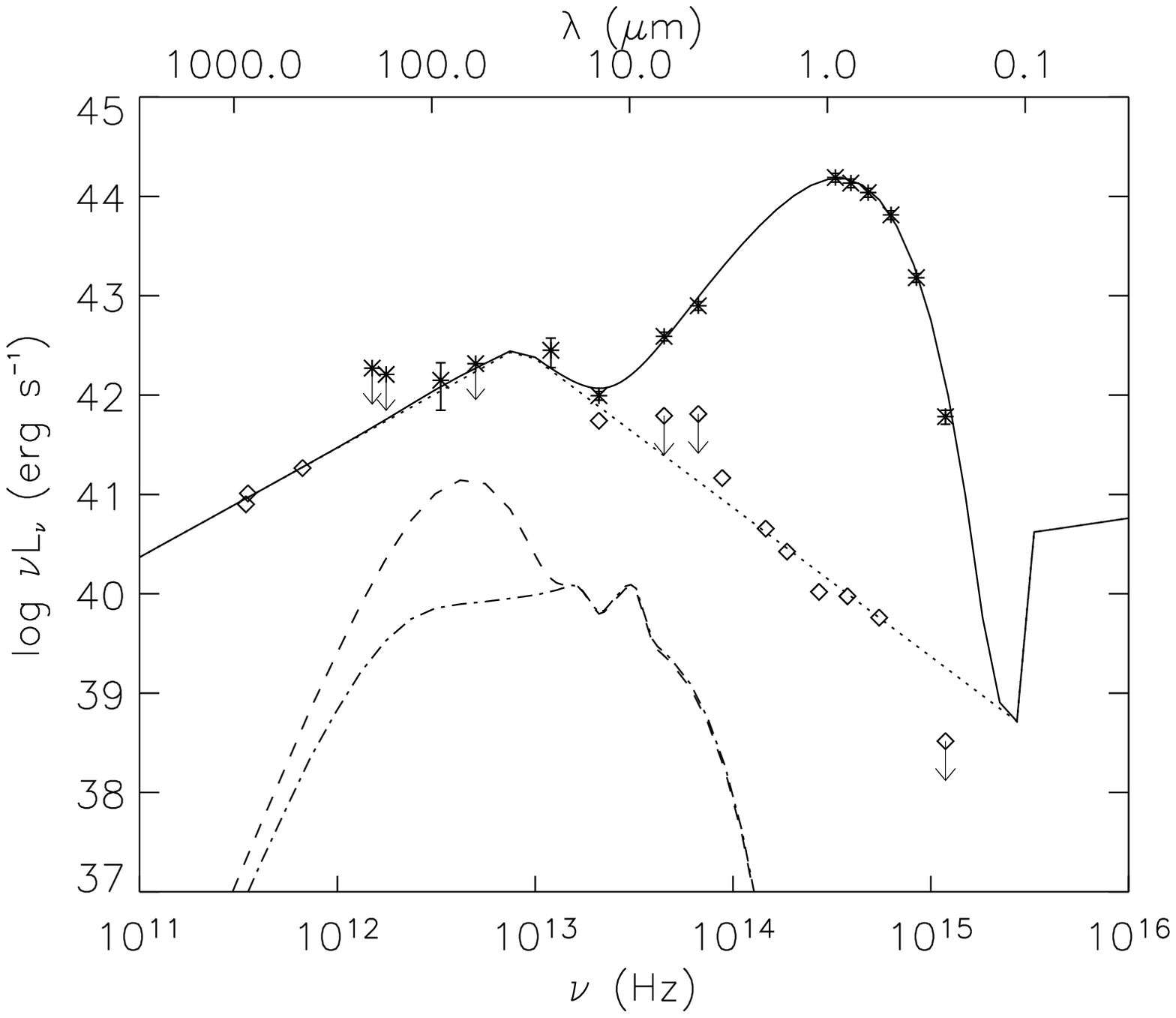}}
\resizebox{8.cm}{!}{\hspace{-1.5cm}\includegraphics{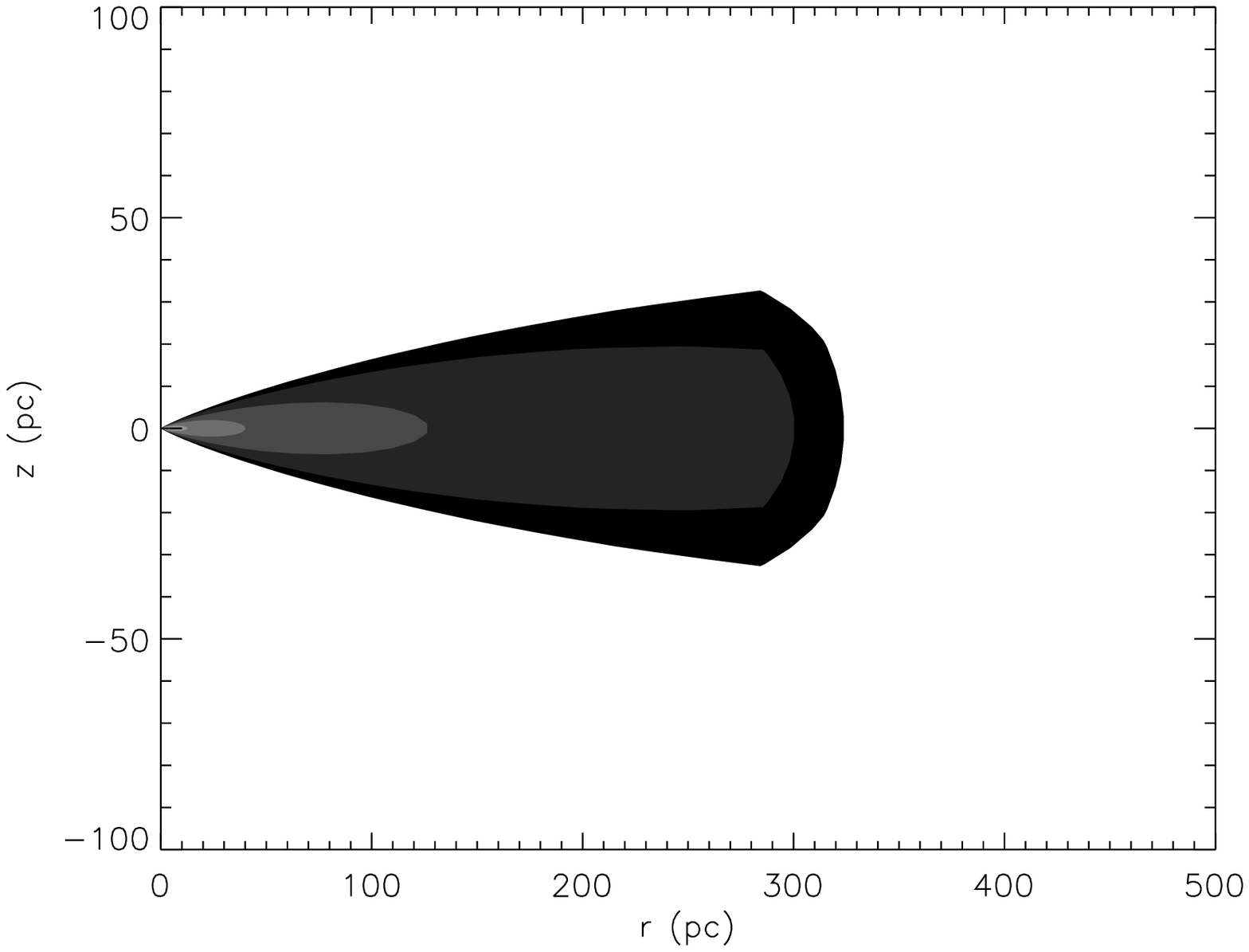}}
\caption{\label{refmodel} {\it Left :} Reference model for \obj. 
The total SED and core SED are
shown as asterisks and diamonds respectively. Our total model SED 
is the solid line. The dashed line is the model output from dusty disk 
The dash-dotted line is the disk emission without the stellar heating,
and the dotted line is the nuclear input spectrum. The model core spectrum
is similar to the input spectrum.
Error bars are omitted where the error is smaller than the symbol.\newline
{\it Right :} The density structure of the disk. Lighter areas represent
denser regions. The disk is axis-symmetric around the $z$-axis.}
\end{figure*}

\section{Analysis and modeling of the SED}
In the total flux SED we detect the large scale radio emission and a
stellar peak of with black body temperature of $4180\pm40$\,K. Both decline
steeply towards the infrared and cannot responsible for
the far-infrared emission.
The core SED shows a typical double power-law shape with a turnover
around 30\um, as expected 
\citep{fal04}. Including only high resolution data, we find spectral 
indices of 0.2 and --2.5 for the long and short wavelenght sides 
respectively. Including lower resolution VLA data results in a spectral
index of --0.1, while including the sub-mm data results in 0.1. 
These values are typical for FR\,I synchrotron cores.

The 90\um\ detection is well fitted by the power-law, and due to the
large calibration error is fully consistent with the range of spectral
indices we find. Therefore we conclude that non-thermal emission is
the dominant source for the far-infrared, and find the same to be
true for the sub-millimeter. The 25\um\ detection is consistent with
both power-laws fit, but due to the large error margin we cannot
fully exclude a more significant thermal contribution at this wavelength.
Our fit results in a sharp and probably unphysical turn-over, but fitting
a more physical curved spectrum does not alter our findings (see below).

This implies that the nuclear disk cannot contribute more than 50\%
of the infrared flux. We have tested this with Monte Carlo radiative 
transfer models, using a setup identical to \citet{bd03} and disk 
parameters from observations by \citet{jaf96} and \citet{mart00}. 
The dust is heated by both the stellar radiation and
the active nucleus. We add a UV--X-ray component to the observed core
spectrum which maximizes the heating from the nucleus. The disk mass,
mass distribution and geometry are varied within the constraints 
from observations. From our models
we find that the disk never contributes more than 
10\% of the infrared emission (see Fig.~\ref{refmodel}). We also
find that stars dominate the heating of cool dust, while the AGN
dominates heating within a 10\,pc radius. However, none of our models 
provide an explanation for the absence of a UV core. Either there 
is more dust in the line of sight than we account for, or the intrinsic 
core spectrum steepens significantly towards the UV.

\begin{figure*}[!Ht]
\resizebox{8.cm}{!}{\hspace{-1.5cm}\includegraphics{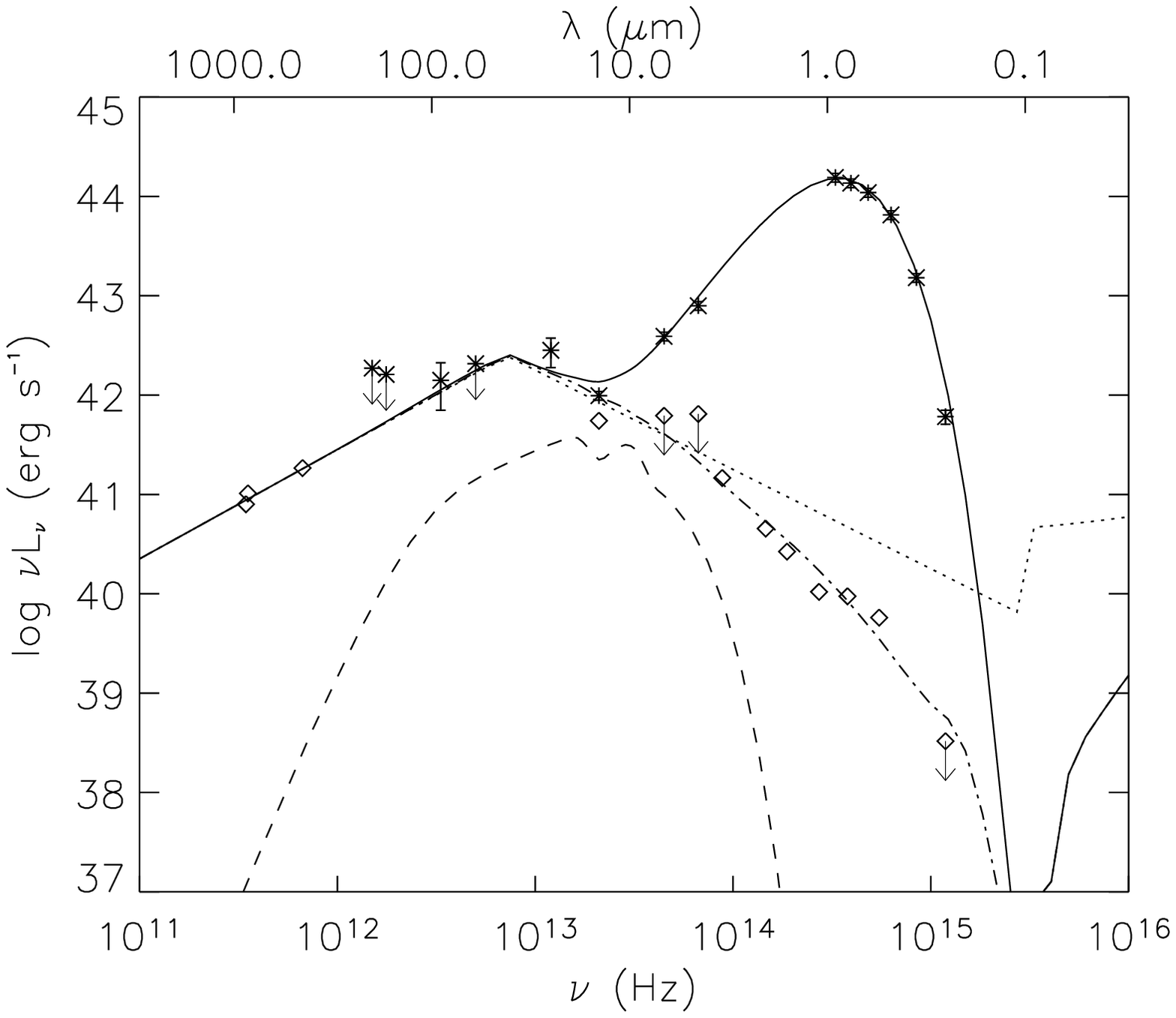}}
\resizebox{8.cm}{!}{\hspace{-1.5cm}\includegraphics{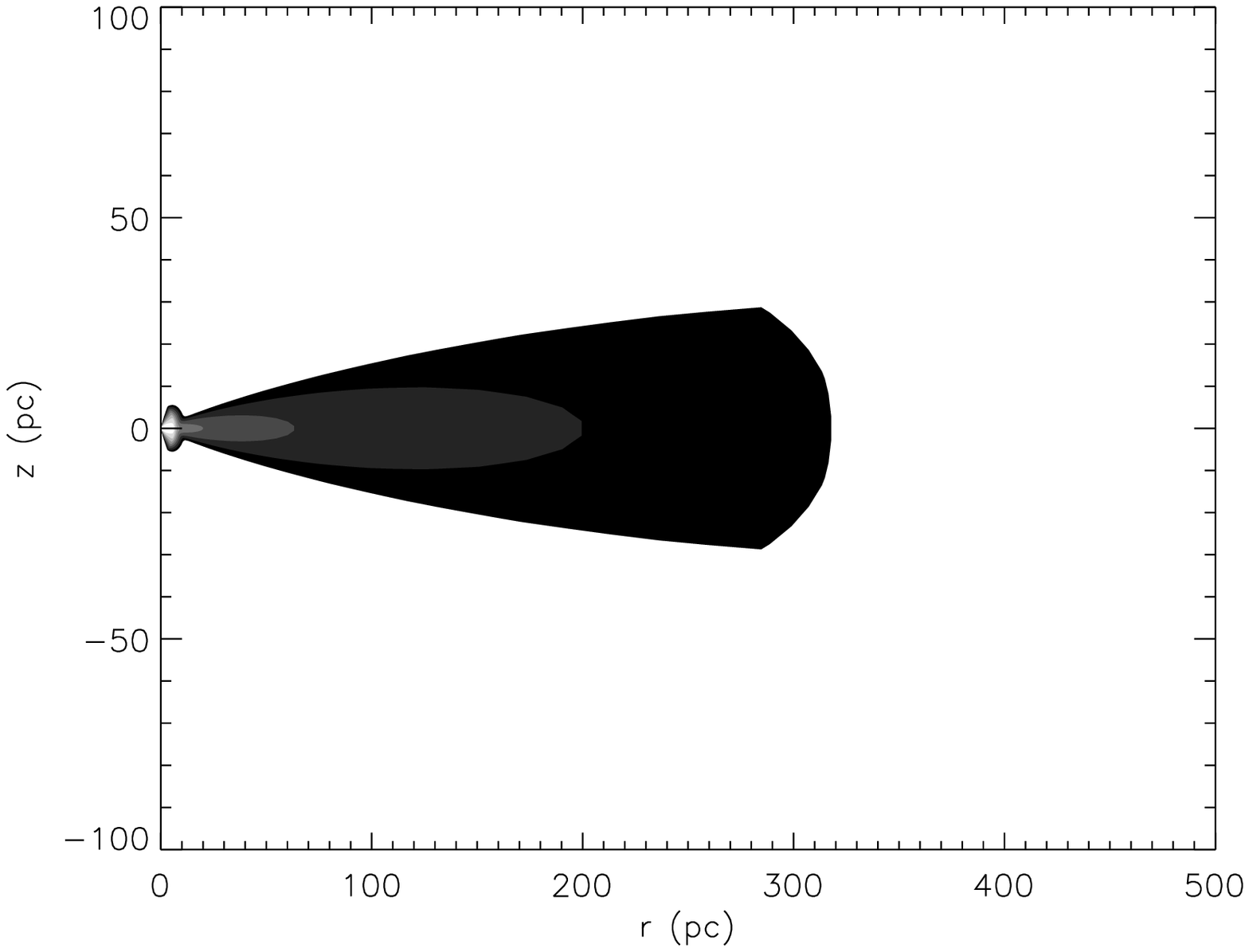}}
\caption{\label{density} As Fig.~\ref{refmodel}, but for the disk with inflated
inner region. Here the dash-dotted line is the model core spectrum. The
nuclear input spectrum is flattened in the optical range to improve
the fit.}
\end{figure*}

Within the highly constrained observations, more dust can only be present
in the innermost region of the nuclear disk, which is unresolved. Evidence
for a thicker disk within 6\,pc comes from detections of HI absorption
\citet{lang00}. We constructed model disks which have an inflated inner
region, with parameters identical to the HI disk (see Fig.~\ref{density}. 
As expected, these
models boost the 2--20\um\ emission by an order of magnitude, while
the far-infrared is unaffected. The inner regions absorb the core from
the infrared to the UV, but by flattening the intrinsic core spectral 
index in that range, the fit to the observed core spectrum improves and 
accounts for the lack of a UV core.

Alternatively, if FR\,Is are mis-aligned BL~Lacs, they may have intrinsically
curved core spectra. Fitting the core spectrum with a model from \citet{cg99}
we can indeed explain the absence of a UV core with a BL~Lac type spectrum. 
All model parameters fall well within the observed range for BL~Lacs, only 
the beaming parameter is adjusted for the larger inclination.

Since the observations span about 2 decades in time, core variability can be
significant. We therefore allow a factor 2 in our fits, which makes it
impossible to distinguish between either solution. The 25 and 90\um\ 
detection are also consistent with both, although there is marginally
more room for thermal 25\um\ emission with the curved spectrum. 
However, the curved core spectrum provides much less heating power to
the disk compared to our models. Therefore, the disk would be less
bright, making it difficult to account for any thermal emission in
the MFIR, and enforcing the conclusion that non-thermal emission is 
the dominant MFIR process.

\section{Consequences for unification models}
In the FR\,II unification models, the role of the hypothetical torus
is to obscure the nucleus at larger inclinations. We have shown that
under certain assumptions, the inner parts of the nuclear disk are 
inflated and do exactly that. Classical tori in FR\,IIs require
$\tau_{\rm V} \gg 1$, whilst the disk in \obj\ has $\tau_{\rm V} \sim 1$,
so in that sense, it is not an obscuring torus. Perhaps we
need to adjust our definition and allow tori to have a range
of optical depths. This would imply that some FR\,Is have a torus,
although it is optically thin, and the inner region of the nuclear
disk in \obj\ would potentially be one. 

If FR\,Is indeed have thin tori, there may no longer be need for
two separate unification models. Instead, there may be a continuous
distribution of torus optical depths, in which FR\,Is and IIs 
predominantly populate different regions.
However, there is not necessarly a sharp 
divide. The SED of the nuclear disk is in good agreement with the
infrared emission found in broad-line radio galaxies \citep{ivb00}. 
Assuming their mid-infrared emission is thermal, this would imply that
some FR\,IIs could also have thin tori.

Finally, we would like to stress the importance of using the highest
possible resolution data to determine the presence of a thermal
infrared bump. Lower resolution radio data include a significant
contribution from older electron populations, which flatten the
spectrum and thus underpredict the non-thermal infrared emission.
In \obj\ the extrapolation from the total radio spectrum would
predict a significant thermal infrared bump, which is obviously
not the case. In addition, we add to other evidence that the 
far-infrared emission is dominated by stellar heating, which 
makes it highly unsuitable for testing unification models.
The best wavelength to detect thermal 
emission from a torus would be between 10 and 20\um, where both 
the non-thermal core and the stellar emission sharply decline, and 
the torus emission peaks. 


\begin{theacknowledgments}
Thanks to Mark Birkinshaw for providing access to not yet published
ISO CAM results.
IMvB acknowledges ESA for a Research Fellowship, and thanks the 
Space Telescope Science Institute DDRF and the Van Leeuwen Boomkamp 
Fonds from the KNVWS for additional financial support.
\end{theacknowledgments}


\bibliographystyle{aipproc}   


\begin{thebibliography}{9}
\expandafter\ifx\csname natexlab\endcsname\relax\def\natexlab#1{#1}\fi
\providecommand{\enquote}[1]{``#1''}
\expandafter\ifx\csname url\endcsname\relax
  \def\url#1{\texttt{#1}}\fi
\expandafter\ifx\csname urlprefix\endcsname\relax\def\urlprefix{URL }\fi

\bibitem[{Barthel}(1989)]{pdb89}
{Barthel}, P.~D., \emph{ApJ}, \textbf{336}, 606 (1989).

\bibitem[{Urry} and {Padovani}(1995)]{up95}
{Urry}, C.~M., and {Padovani}, P., \emph{PASP}, \textbf{107}, 803 (1995).

\bibitem[{Falcke} et~al.(2004)]{fal04}
{Falcke}, H., {K{\" o}rding}, E., and {Markoff}, S., \emph{A\&A}, \textbf{414},
  895--903 (2004).

\bibitem[{van Bemmel} and {Dullemond}(2003)]{bd03}
{van Bemmel}, I.~M., and {Dullemond}, C.~P., \emph{A\&A}, \textbf{404}, 1
  (2003).

\bibitem[{Jaffe} et~al.(1996)]{jaf96}
{Jaffe}, W., {Ford}, H., {Ferrarese}, L., F., v., and {O'Connell}, R.~W.,
  \emph{ApJ}, p. 214 (1996).

\bibitem[{Martel} et~al.(2000)]{mart00}
{Martel}, A.~R., {Turner}, N.~J., {Sparks}, W.~B., and {Baum}, S.~A.,
  \emph{ApJS}, \textbf{130}, 267--338 (2000).

\bibitem[{van Langevelde} et~al.(2000)]{lang00}
{van Langevelde}, H.~J., {Pihlstr{\" o}m}, Y.~M., {Conway}, J.~E., {Jaffe}, W.,
  and {Schilizzi}, R.~T., \emph{\aap}, \textbf{354}, L45--L48 (2000).

\bibitem[{Chiaberge} and {Ghisellini}(1999)]{cg99}
{Chiaberge}, M., and {Ghisellini}, G., \emph{\mnras}, \textbf{306}, 551--560
  (1999).

\bibitem[{van Bemmel} et~al.(2000)]{ivb00}
{van Bemmel}, I.~M., {Barthel}, P.~D., and {de Graauw}, T., \emph{A\&A},
  \textbf{359}, 523--530 (2000).

\end{thebibliography}

\end{document}